\documentclass[12pt,preprint]{aastex}

\usepackage{emulateapj5}

\def\hii{H\,{\sc ii}}
\def\hi{H\,{\sc i}}
\def\kms{\relax \ifmmode {\,\rm km~s}^{-1}\else \,km~s$^{-1}$\fi}
\def\cm-3{\relax \ifmmode {\,\rm cm}^{-3}\else \,cm$^{-3}$\fi}
\def\Jb{\relax \ifmmode {\,\rm Jy\,beam}^{-1}\else \,Jy\,beam$^{-1}$\fi}
\def\mJb{\relax \ifmmode {\,\rm mJy\,beam}^{-1}\else \,mJy\,beam$^{-1}$\fi}
\def\deg{\hbox{$^\circ$}}
\def\arcmin{\hbox{$^\prime$}}
\def\arcsec{\hbox{$^{\prime\prime}$}}
\def\secd#1.#2{ #1\farcs#2 }               
\def\e{$\pm$}
\def\x{$\times$}
\def\j21{$J$=2$\rightarrow$1}

\begin{document}

\title{Interferometric Mapping of Magnetic Fields in Star-forming Regions II. NGC\,2024 FIR\,5}
\author{Shih-Ping Lai\altaffilmark{1}, Richard M. Crutcher, 
Jos\'e M. Girart\altaffilmark{2}, and Ramprasad Rao\altaffilmark{3}}
\affil{Astronomy Department, University of Illinois, 1002 W. Green Street, Urbana, IL 61801; \\
slai@thisvi.jpl.nasa.gov, crutcher@astro.uiuc.edu, jgirart@am.ub.es, ramp@oddjob.uchicago.edu}
 
\altaffiltext{1}{Current address: Jet Propulsion Laboratory, California Institute of Technology, MS 169-506, Pasadena, CA 91109}
\altaffiltext{2}{Current address: Departament d'Astronomia i Meteorologia, Universitat de Barcelona, 08028 Barcelona, Catalunya, Spain}
\altaffiltext{3}{Current address: Department of Physics, University of Chicago}

\begin{abstract}
We present the first interferometric polarization maps of the NGC\,2024
FIR\,5 molecular core obtained with the BIMA array at approximately 
2\arcsec\ resolution.
We measure an average position angle of $-$60\deg\e6\deg\ in the main core 
of FIR\,5 and 54\deg\e9\deg\ in the eastern wing of FIR\,5.
The morphology of the polarization angles in the main core of FIR\,5 
suggests that the field lines are parabolic with a symmetry axis
approximately parallel to the major axis of the putative disk in FIR\,5,
which is consistent with the theoretical scenario that
the gravitational collapse pulled the field lines into an hour-glass shape.
The polarization percentage decreases toward regions with high intensity
and close to the center of the core, suggesting that the dust
alignment efficiency may decrease at high density.
The plane-of-sky field strength can be estimated with 
the modified Chandrasekhar-Fermi formula, and the small dispersion
of the polarization angles in FIR\,5 suggests that the magnetic field
is strong ($\gtrsim 2$ mG) and perhaps dominates the turbulent motions
in the core.

\end{abstract}

\keywords{ISM: magnetic fields -- ISM: individual: NGC 2024 -- polarization-- star: formation -- techniques: interferometric }

\section{Introduction}

Magnetic fields are thought to play a significant role 
in all stages of star formation 
(e.g., recent reviews by Mouschovias \& Ciolek 1999; Shu et al.\ 1999).
However, the magnetic field is the most poorly measured quantity
in the star formation process.    Observations of the linear polarization 
from the thermal emission of magnetically aligned dust grains provide 
a relatively easy approach to explore the magnetic field morphology
(Heiles et al.\ 1993).   Such observations give the field direction 
in the plane of the sky perpendicular to the direction of polarization
(Davis \& Greenstein 1951; Roberge 1996).

Information on the magnetic field morphology is useful
for testing the predictions of theoretical models and simulations.
The standard star formation theory predicts certain
morphological evolution of magnetic fields.  Theory predicts that 
molecular clouds will tend to be flat with their minor axes parallel to
the field lines, because magnetic fields prevent collapse
perpendicular to the direction of the field lines (Mouschovias 1976).
As contracting cores form, the field morphology achieves an ``hourglass"
shape with a collapsing accretion disk $\sim$100 AU at the ``pinch"
and a magnetically supported envelope $\sim$1000 AU
(Fiedler \& Mouschovias 1993; Galli \& Shu 1993).
Furthermore, the rotation of disks may twist field lines
into the direction along the disk and form a toroidal morphology
(Holland et al.\ 1996).
Numerical simulations performed by Ostriker, Gammie, \& Stone (1999)
show that the field morphology is more random
for larger ratios of the thermal to magnetic energy.

Because of the high sensitivity requirement in polarization observations, 
most previous observations of even the nearest star-forming regions have 
been made with single-dish telescopes whose large beams cover a region 
greater than the physical extent of protostellar cores.
For example, the angular resolution of the James Clerk Maxwell Telescope 
(JCMT) at 850 $\mu$m is $\sim$14\arcsec, which corresponds to $\sim$6000 AU 
at the distance of the Orion Molecular Cloud.
Therefore, in order to test theoretical models and simulations of star 
formation, 
it is essential to obtain high-resolution observations of magnetic fields.   
The best available approach to acquire high resolution 
information is to conduct interferometric observations.
Pioneer interferometric polarization observations at millimeter wavelengths
have been done with the Owens Valley Radio Observatory 
(Akeson \& Carlstrom 1997).
Recently, the Berkeley-Illinois-Maryland-Association millimeter array
(BIMA) has been able to successfully provide extended 
polarization maps with high resolution up to 2\arcsec\ 
(Rao et al.\ 1998; Girart, Crutcher, \& Rao 1999; 
Lai et al.\ 2001, hereafter Paper I).

NGC\,2024, a massive star formation region in the Orion B Giant Molecular 
Cloud  (distance $\sim$ 415 pc: Anthony-Twarog 1982), is a good
candidate to explore the magnetic field structure at small 
scales.   It contains a luminous \hii\ region with a north-south 
molecular ridge at its center corresponding to the dust lane in the optical 
image.   
Mezger et al.\ (1988, 1992) identify seven dust cores
in the molecular ridge at 1300 and 350 $\mu$m 
and assign designations using the acronym 'FIR'.
They interpret these cores as
isothermal protostars in the stage of free-fall contraction.
However, detailed studies of FIR\,4/5/6 show that these cores contain 
either a near-infrared source or strong outflows that are traditionally 
related to protostellar cores in a more evolved stage 
(Moore \& Chandler 1989; Chandler \& Carlstrom 1996).
Our interest here focuses on the brightest core, FIR\,5. 
With a highly collimated unipolar molecular outflow extended 
over $\sim$5\arcmin\ south of the core (Richer et al.\ 1992), FIR\,5 
appears to be the most evolved object among the FIR cores in NGC\,2024.
Continuum observations at 3 mm by Wiesemeyer et al.\ (1997)
resolve FIR\,5 into two compact cores, and they suggest FIR\,5 
is a binary disk with an envelope.
The magnetic field structure in NGC\,2024 has been mapped with 
VLA OH and \hi\ Zeeman observations by Crutcher et al.\ (1999)
and with far-infrared dust polarization by Dotson at al.\ (2000).
Both observations have beam sizes larger than the FIR\,5 core.
Our observations provide new information 
on detailed field morphology in the FIR\,5 core.

\section{Observations and Data Reduction}

The observations were carried out from 1999 March to 2001 February
using nine BIMA antennas with 1-mm Superconductor-Insulator-Superconductor
(SIS) receivers and quarter-wave plates.  
The digital correlator was set up to observe both continuum and the
CO \j21\ line simultaneously.  The continuum was observed
with a 750 MHz window centered at 226.9 GHz in the lower sideband
and a 700 MHz window centered at 230.9 GHz in the upper sideband.
Strong CO $J$=2--1 line emission was isolated in
an additional 50 MHz window in the upper sideband.
The primary beam was $\sim$50\arcsec\ at 1.3 mm wavelength.    
Data were obtained in the D, C, and B array configurations,
and the projected baseline ranges were 4.5--20, 5--68, 
and 5--180 kilowavelengths.
The integration time in the D, C, and B array was 3.4, 19.1, and 11 hours,
respectively.
The D-array observations were made first on 1999 March with a pointing
center between NGC\,2024 FIR\,5 and FIR\,6 ($\alpha_{2000}=5^h41^m45\fs0$,
$\delta_{2000}=-1\deg55\arcmin55\farcs0$).  Because the D-array data 
showed significant amount of polarized flux toward FIR\,5 near the edge
of the primary beam, the follow-up C- and B-array observations were 
therefore made with a pointing center at FIR\,5 
($\alpha_{2000}=5^h41^m44\fs35$, $\delta_{2000}=-1\deg55\arcmin42\farcs0$).

The BIMA polarimeter and the calibration procedure are described 
in detail in Paper I (also see Rao et al.\ 2001, in preparation).
The average instrumental polarization of each antenna 
was 5.6\%\e0.4\% for our observations.  
The Stokes $I$ image of the continuum was made by mosaicing 
the B-, C-, and D-array data
with Briggs' robust weighting of 0.5 (Briggs 1995; Sault \& Killeen 1998) 
to obtain a smaller synthesized beam (HPBW 2\farcs3\x1\farcs4, PA=9\deg)
without losing significant amount of flux.  D-array data were needed
to recover the extended emission in order to obtain better measurements
of the polarization percentage.  The Stokes $Q$ and $U$ images were made with 
B- and C-array data only, because the precise pattern of the off-axis 
polarization across the primary beam is unknown.   Rao et al.\ (1999)
measured the off-axis instrumental polarization at selected positions and 
showed that it only provides an uncertainty less than 0.5\% (comparable to 
the uncertainty of the leakages) within 4/5 of the primary beam; therefore, 
the off-axis polarization calibration was ignored.  Natural weighting was 
used to produce the maps of the Stokes $Q$ and $U$ in order 
to obtain the highest  S/N ratio, and the resulting synthesized beam was 
2\farcs4$\times$1\farcs4 with PA=8\deg.  

Maps of Stokes $Q$ and $U$ were deconvolved and binned to approximately 
half-beamwidth per pixel (1\farcs2\x0\farcs6) to reduce oversampling 
in our statistics.   These maps were
combined to obtain the linearly polarized intensity ($I_p$), the position
angle ($\phi$), and the polarization percentage ($p$),
along with their uncertainties as described in Section 2 of Paper I. 
When weighted with $I_p$, the average measurement uncertainty in 
the position angle for our observations was 6\fdg0\e2\fdg3.

\section{Results and Analysis}

Figure 1 shows the B+C-array Stokes I map (contours) superposed on 
the D-array Stokes I map (grey-scale).  
The B+C-array map resolves FIR\,5 and FIR\,6 into several clumps 
with the highest resolution ever obtained (1\farcs6\x1\farcs2, PA=11\deg).   
The clumps with peak flux stronger than 7$\sigma_I$ ($1\sigma_I$ = 3.3 \mJb) 
are named and tabulated in Table 1.  
Figure 2 displays the mosaiced map of NGC\,2024 FIR\,5 
overlaid with polarization vectors. 
This map contains B, C, and D array data.
Polarization vectors are plotted 
at positions where the observed linearly polarized intensity is 
greater than 3$\sigma_{I_p}$ (1$\sigma_{I_p}$ = 2.1 \mJb) and 
the total intensity is greater than 3$\sigma_I$ (1$\sigma_I$ = 7.4 \mJb; 
note that $\sigma_I$ is dominated by incomplete deconvolution rather than 
thermal noise).   
Under these criteria, the polarized emission extends over an area of $\sim$8 
beam sizes.  Table 2 lists the polarization measurements in NGC\,2024 FIR\,5 
at selected positions separated by approximately the synthesized beamsize.   
The distributions and the model fitting of the polarization angle 
are plotted in Figure 3 and Figure 4 respectively, and 
the distribution of the polarization percentage is shown in Figure 5.

\subsection{Continuum and Polarized Emission}

The continuum emission of FIR\,5 can be separated into three main
components -- FIR\,5:Main, FIR\,5:NE, and FIR\,5:SW (Figure 1).  
FIR\,5:Main are resolved into seven clumps, and the seven clumps 
are assigned designations with 'LCGR' (Table 1).  Only
the two brightest clumps (FIR\,5:LCGR 4 and 6) were previously
identified at 96 GHz (Wiesemeyer et al.\ 1997; 
note that we have renamed these two clumps).   
Wiesemeyer et al.\ (1997) suggested that FIR\,5 is a binary disk; 
however, our high resolution data present complex morphology in FIR\,5.
Our data also show that the continuum of FIR\,6 
consists of two compact sources.  The three main components of FIR\,5 
and the two compact sources of FIR\,6 are clearly seen in
the 450 $\mu$m continuum maps of Visser et al.\ (1998),
suggesting that they are dust dominated sources.

Figure 2 shows that our detection of the polarized flux in FIR\,5 
is mainly distributed in a $\sim$10\arcsec\x4\arcsec\ strip
along the major axis of FIR\,5:LCGR 4 extending into FIR\,5:LCGR 2, 
which is roughly perpendicular to the disk previously proposed by
Wiesemeyer et al.\ (1997).
The peak of the polarized flux occurs at a position about 
1\farcs8 north of the continuum peak between FIR\,5:LCGR 3 and 4.
Detections with $I_p>3\sigma_{I_p}$ were also made toward positions 
near FIR\,5:LCGR 7 and in FIR\,5:NE.

\subsection{Polarization Angle Distribution} 

The histogram of polarization angles in FIR\,5 is shown in Figure 3. 
Most of the polarization detected is associated with FIR\,5:LCGR 4. 
Toward this source the average position angle is $-$60\deg\e6\deg, 
and the angles appear to decrease from south to north from $-$7\deg\ to 
$-$71\deg. 
Compact polarization is also observed associated with FIR\,5:NE and 
FIR\,5:LCGR 7.  Both these sources 
show a similar position angle (average = 54\deg\e9\deg) which is 
very distinct from that of the vectors in FIR\,5:LCGR 4.
Our results are consistent with recent JCMT observations at 850 $\mu$m,
which also show $\phi\sim$ 50\deg\ in FIR\,5:NE, low polarization
around FIR\,5:LCGR 6, and $\phi\sim$ $-$60\deg\ in FIR\,5:LCGR 4
(Matthews et al.\ 2001, private communication).

Since the magnetic field direction inferred from the dust polarization 
is perpendicular to the polarization vectors,  the variation of 
the polarization angles around FIR\,5:LCGR 4 suggests that 
the magnetic field lines are curved.   The curved field lines can be 
successfully modeled with
a set of parabolas with the same focal point, and the best model 
with minimum $\chi^2$ is presented in Figure 4.    The symmetry
axis of the best model is at $-$77\deg, which is consistent with the 
position angle of the line connecting FIR\,5: d and f at $-$67\deg.
The histogram in Figure 4 shows the distribution of the deviation 
between our measurements and the best model, which shows a Gaussian-like
distribution with a dispersion of $\delta\phi_{obs}$=9\fdg2\e3\fdg3.  
After deconvolving the measurement uncertainty in this region
($\sigma_{\phi}=5\fdg8\pm2\fdg2$) from the observed dispersion,
we obtain the intrinsic dispersion of the polarization 
angles $\delta\phi=7\fdg1\pm2\fdg5$.

\subsection{Polarization Percentage Distribution}  

The average polarization percentage is 6.0\%\e1.5\% in the main core 
and 10.5\%\e4.7\% toward the eastern side of FIR\,5.
The polarization percentage distribution is plotted in Figure 5. 
These plots show that the polarization percentage decreases toward 
regions of high intensity as well as toward the cloud center.  
Based on the apparent close
correlation in both Figure 5(a) and 5(b),  we perform a least-squares
fit on $\log_{10}p$ versus $\log_{10}I$ and $p$ versus the
distance from the peak of FIR\,5:LCGR 4 ($R$).  We obtain the following results:
(1) for all data points, $\log_{10}p=(-1.88\pm0.01)-(0.83\pm0.01)
\times\log_{10}I$ with a correlation coefficient of $-$0.86, and
(2) in the main core, $p=(3.60\pm0.63)\%+(1.45\pm0.16)R$ with a correlation 
coefficient of 0.97.
Because higher intensity and smaller radius both imply higher density, 
our results suggest that the polarization percentage decreases toward 
the high density region.    This conclusion is consistent with what we 
reported in Paper I for the W51 e1/e2 cores, and our interpretation
was the decrease of the dust alignment efficiency toward high 
density regions.

\section{Discussion}

\subsection{Magnetic Field and Molecular Core Morphology}

In \S3.2, we showed that the field lines in NGC\,2024 FIR\,5
can be modeled with a set of parabolas which may represent
part of an hour-glass shape.  The reason that the rest of 
the hour-glass shape is not detected could be due to 
the relatively lower dust column density 
to the east of the region where polarization has been detected.
The hour-glass geometry has been predicted by theoretical work 
and simulations (Fiedler \& Mouschovias 1993; Galli \& Shu 1993).
When enough matter collapses along field lines 
and forms a disk-like structure, 
the gravity in the direction along the major axis of the disk
would eventually overcome the supporting forces 
and pull the field lines into an hour-glass shape.
Observations of magnetic fields have found several cases consistent
with this geometry, such as in W3 (Roberts et al.\ 1993; 
Greaves, Murray, \& Holland 1994), OMC-1 (Schleuning 1998), and 
NGC\,1333 IRAS\,4A (Girart, Crutcher, \& Rao 1999).
Our results in NGC\, 2024 FIR\,5 provide a possible supporting example 
to the theoretical models at a small scale of few thousand AU.
This scale is much smaller than the hour-glass shapes observed 
in W3 and OMC-1 ($10^4-10^5$ AU), and is comparable to that 
in NGC\,1333 IRAS\,4A.

The symmetry axis of our hour-glass model is approximately parallel to 
the binary disk previously proposed by Wiesemeyer et al.\ (1997).
Such a coincidence could be an indirect evidence for the existence 
of a collapsing disk.   Although our high resolution map shows that 
FIR\,5:Main is more complicated than a binary disk, 
it is still possible that there is an east-west disk consisting of 
four clumps (FIR\,5:LCGR 1, 4, 6, and 7), with
the remaining three clumps either collapsing into the disk
or being ejected through outflows.  
FIR\,5:LCGR 2 is at $\sim-$145\deg\ direction with respect to FIR\,5:LCGR 4,
which is close to the unipolar outflow at $\sim-$170\deg. 
Since outflows could be deflected by magnetic fields
away from the protostellar cores (Girart, Crutcher, \& Rao 1999), 
it is possible that FIR\,5:LCGR 2 traces the direction of the outflow
near the central core.    
Further high resolution kinematic study is needed to examine 
the above speculation.

\subsection{Estimation of the Magnetic Field Strength}

Although the magnetic field strength cannot be directly inferred from
polarization of dust emission, Ostriker, Stone, \& Gammie (2001) show
that the Chandrasekhar-Fermi formula (Chandrasekhar \& Fermi 1953) 
modified with a factor of 0.5 can provide accurate estimates
of the plane-of-sky field strength under strong field cases
($\delta\phi\leq25\deg$) for their simulations of magnetic turbulent 
clouds.  Therefore, the projected field strength ($B_p$) 
can be expressed as
\begin{equation}
B_p = 0.5 \sqrt{4\pi\bar{\rho}}~~\frac{\delta v_{los}}{\delta\phi}
= 8.5~\frac{\sqrt{n_{H_2}/(10^6\cm-3)}~\Delta v/\kms}
{\delta\phi/1\deg}~~{\rm mG},
\end{equation}
\noindent where $\bar{\rho}$ is the average density, $\delta v_{los}$
is the rms line-of-sight velocity, the number density
of molecular hydrogen $n_{H_2} = 2.33~m_H n_H$, and the linewidth
$\Delta v=\sqrt{8\ln2}~\delta v_{los}$.  

To estimate $B_p$ in NGC\,2024 FIR\,5, we must first carefully
determine the density of the dust core and the turbulent linewidth.  
Mezger et al.\ (1992) derived high density ($n_H\sim10^8$\cm-3) 
in FIR 1-7 from dust emission, but their value is significantly
higher than that derived from molecular studies
($\sim$10$^6$\cm-3 from CS: Schulz et al.\ 1991; 
 $\sim$10$^7$\cm-3 from C$^{18}$O: Wilson, Mehringer, \& Dickel 1995).
It is possible that depletion of molecules in the dense core
causes molecules to only trace the envelope of the dense core;
on the other hand, Chandler \& Carlstrom (1996) showed
that the kinetic temperature in the FIR cores is higher than what
Mezger et al.\ (1992) assume, implying a lower density.
Mangum, Wootten, \& Barsony (1999) studied the kinetic temperature 
in NGC\,2024 with multi-line observations of formaldehyde (H$_2$CO), 
and their results supported the arguments of Chandler \& Carlstrom (1996).  
We therefore adopt $n_{H_2}\sim2\times10^6$\cm-3 for FIR\,5 from Mangum, 
Wootten, \& Barsony's large velocity gradient (LVG) calculation.   
We also adopt a weighted linewidth of $\Delta v=2.04\pm0.03$\kms\ 
at the position at the peak of FIR\,5 from Mangum, Wootten, 
\& Barsony (1999). Note that, just as for other molecules, 
the peak of H$_2$CO emission does not coincide with the continuum peak;
therefore, the H$_2$CO emission may come from the envelope of the FIR\,5.
This may be an advantage because the H$_2$CO linewidth is less likely
contaminated by the possible dynamical motions in the core and 
better represents the turbulent motion.

In \S3.2, we calculated the angle dispersion $\delta\phi=7\fdg1\pm2\fdg5$
after taking out the systematic field structure in order to identify
the dispersion purely from the Alfv\'enic motion. 
Along with the parameters discussed in the previous paragraph,
we obtain $B_p\sim$3.5 mG.  We can also calculate the lower limit
of $B_p$ using the largest angle dispersion, which is the dispersion
before taking out the parabolic model ($\delta\phi=$13\fdg1\e4\fdg6); thus,
$B_{p,min}\sim$1.9 mG.     Our estimate of the
plane-of-sky field strength is much larger than the line-of-sight 
field strength measured by Crutcher et al.\ (1999) from OH Zeeman
observation, which is $\sim65\mu$G at FIR\,5. 
Except in the unlikely case that the magnetic field direction lies 
almost on the plane of the sky, the low field strength detected 
with the Zeeman measurement can be explained by 
(1) the beam-averaging over small scale field structure, and/or
(2) that OH line does not trace high density region of the dust cores.

\subsection{Turbulent Energy vs.\ Magnetic Energy}

The small dispersion of the polarization angles observed by us
may imply that the turbulent motion 
is not strong enough to disturb the magnetic field structure.
In order to quantitatively discuss the the relative importance 
of the turbulent and magnetic energy in NGC\,2024 FIR\,5,
we calculate the ratio of the turbulent to 
magnetic energy 
\begin{equation}
\beta_{turb}\equiv\frac{\sigma_{turb}^2}{V_A^2}, 
\end{equation}
where $\sigma_{turb}$ is the turbulent linewidth and 
$V_A=|\bf{B}|/\sqrt{4\pi\bar{\rho}}$ is the Alfv\'en speed.  
Statistically, $<B_p^2>=\frac{2}{3}~|\bf{B}|^2$, thus 
$V_A$ can be estimated from Eq.\ (1).
Therefore,
\begin{equation}
\beta_{turb}\approx\frac{8}{3}~\frac{\sigma_{turb}^2}{\delta v_{los}^2}~\delta\phi^2.
\end{equation}
In the general conditions prevalent in molecular clouds, the thermal linewidth 
$\sigma_{thermal}$ is much smaller than the turbulent linewidth, so
$\delta v_{los}^2=\sigma_{turb}^2+\sigma_{thermal}^2\approx\sigma_{turb}^2$.
The ratio of the turbulent to magnetic energy simply depends on
the polarization angle dispersion,
\begin{equation}
\beta_{turb}\approx\frac{8}{3}~\delta\phi^2=8\times10^{-4}(\frac{\delta\phi}{1\deg})^2.
\end{equation}

The 0.5 factor in Eq (1) was obtained under the assumption
that the dust alignment efficiency is uniform throughout the cloud.   
If dust alignment efficiency is lower in the regions with higher density
as we suggest in \S3.3, $\beta_{turb}$ would be larger for 
the same $\delta\phi$ and the correction depends on the degree
of depolarization.   For our case, the most conservative
estimate of the angle dispersion is $\delta\phi<13\fdg1$, 
which leads to a small turbulent to magnetic energy ratio, 
$\beta_{turb}<0.14$. 
Therefore, the magnetic field most likely dominates the turbulent motion
in the core region of NGC\,2024 FIR\,5.
This is also the case in W51 e1/e2 (see Paper I).

\section{Conclusions}

High-resolution polarization observations are needed to explore
the detailed magnetic field structure in star-forming cores.
We have obtained a continuum map of NGC\,2024 FIR\,5 and FIR\,6
and a polarization map of NGC\,2024 FIR\,5 
with the highest resolution ever obtained ($\sim$2\arcsec).  
Our observations resolve FIR\,5 and FIR\,6 into several continuum
clumps.  The information revealed by our polarization observations
of FIR\,5 are summarized below:

\begin{itemize}
\item Extended polarization is detected associated with the main core
at $-$60\deg\e6\deg.  Compact polarization is also
observed toward the eastern side of FIR\,5 at 54\deg\e9\deg.
The polarization is low between these two regions and 
the secondary intensity peak of FIR\,5 has an upper limit of 6\%.

\item The magnetic field lines in the core are systematically curved 
with a symmetry axis close to the major axis of a putative disk.
This is consistent with an hour-glass morphology for the magnetic 
fields predicted by theoretical works.

\item The polarization percentage decreases toward regions with high 
intensity and short distance to the center of the core.  The tight 
correlations imply that the depolarization is a global effect
and may be caused by the decrease of the dust alignment efficiency 
in high density regions.

\item The small dispersion of the polarization angles in the core
suggests that the magnetic field is strong ($\gtrsim$ 2 mG)
and the ratio of the turbulent to magnetic energy is small 
($\beta_{turb}<0.14$).
Therefore, the magnetic field most likely dominates turbulent motions
in NGC\,2024 FIR\,5. 

\end{itemize}

\acknowledgments

This research was supported by NSF grants AST 99-81363 and AST 98-20651.
We would like to thank the staff at Hat Creek, especially Rick Forster
and Mark Warnock for assistance with the polarimeter control system.
We also thank Charles Gammie for his helpful comments.

\clearpage

\begin{table}
\caption{Compact continuum sources in FIR\,5 and FIR\,6}
\vspace*{0.3cm}
\begin{tabular}{lcccc}
\hline
\hline
	& $\alpha_{2000}$ & $\delta_{2000}$ &  $I$ \tablenotemark{a} & $p$   \\
Source	& $^h~~^m~~~^s~~$ & ~\deg~~~\arcmin~~~\arcsec~   & (\mJb)  & (\%)  \\
\hline
FIR\,5:LCGR 1 & 5~41~44.00	& -1~55~40.7  &  46 & $<$ 17  \\
FIR\,5:LCGR 2 & 5~41~44.10   & -1~55~44.0  &  64 & 10.3    \\
FIR\,5:LCGR 3& 5~41~44.23	& -1~55~36.8  &  39 & $<$ 25 \\
FIR\,5:LCGR 4 \tablenotemark{b} & 5~41~44.25   & -1~55~40.8  & 293 & 3.8  \\
FIR\,5:LCGR 5& 5~41~44.32	& -1~55~35.7  &  40 & $<$ 21 \\
FIR\,5:LCGR 6 \tablenotemark{c} & 5~41~44.48	& -1~55~42.2  & 107 & $<$ 6  \\
FIR\,5:LCGR 7& 5~41~44.69	& -1~55~43.5  &  55 & $<$ 14 \\
\hline	                       
FIR\,6 c&  5~41~45.13 	& -1~56~04.2  &  67 &   -- \\
FIR\,6 n&  5~41~45.17 	& -1~56~00.3  &  26 &   -- \\
\hline
\hline
\tablenotetext{a}{The peak intensity is measured from the B+C+D-array map.}
\tablenotetext{b}{FIR\,5-w in Wiesemeyer et al.\ (1997)}
\tablenotetext{c}{FIR\,5-e in Wiesemeyer et al.\ (1997)}
\end{tabular}
\end{table}

\begin{table}
\caption{Polarization Measurements in NGC\,2024 FIR\,5}
\vspace*{0.3cm}
\begin{tabular}{ccccl}
\hline
\hline
Position \tablenotemark{a} & Stokes $I$ \tablenotemark{b} &  Polarization & Polarization & Note \\
(\arcsec,\arcsec) & (\mJb) & Percentage(\%) & Angle (\deg) &\\
\hline
(-0.3,~4.5) & ~26.6\e7.3 & 23.1\e9.5 & -72\e9 & \\
(-1.1,~3.0) & 132.8\e7.3 & 18.7\e1.9 & -71\e2 & Polarization Peak\\
(-1.2,~1.2) & 300.6\e7.3 & ~3.9\e0.7 & -64\e5 & Intensity Peak \\
(-1.8,~4.5) & ~41.9\e7.3 & 19.5\e6.1 & -66\e7 & \\
(-2.6,~1.2) & 153.7\e7.3 & ~4.9\e1.4 & -39\e8 & \\
(-2.6,-0.8) & 137.1\e7.3 & ~6.5\e1.6 & -39\e7 & \\
(-2.6,-2.7) & ~45.6\e7.3 & 15.4\e5.2 & -~8\e9 & \\
(-3.9,-0.9) & ~53.2\e7.3 & 20.0\e4.8 & -42\e6 & \\
(-4.8,-2.4) & ~39.1\e7.3 & 17.3\e6.3 & -39\e9 & \\
(10.2,~8.9) & ~31.7\e7.3 & 27.1\e9.2 & ~63\e7 & FIR\,5:NE \\
(~4.1,-0.5) & ~75.7\e7.3 & ~9.3\e2.9 & ~31\e9 & near FIR\,5:LCGR 7 \\
\hline
-- 	    & -- & ~6.0\e1.5 & -60\e6 & average of the main core\\
--	    & -- & 10.5\e4.7 & ~54\e9 & average of the eastern side of FIR\,5\\
\hline
\hline
\tablenotetext{a}{Offsets are measured with respect to        
the phase center: $\alpha_{2000}$=19$^h$23$^m$44\fs2,
$\delta_{2000}$=14\deg30\arcmin33\farcs4.}
\tablenotetext{b}{Stokes $I$ is measured from the B+C+D-array map.}
\end{tabular}
\end{table}

\begin{figure}
\plotone{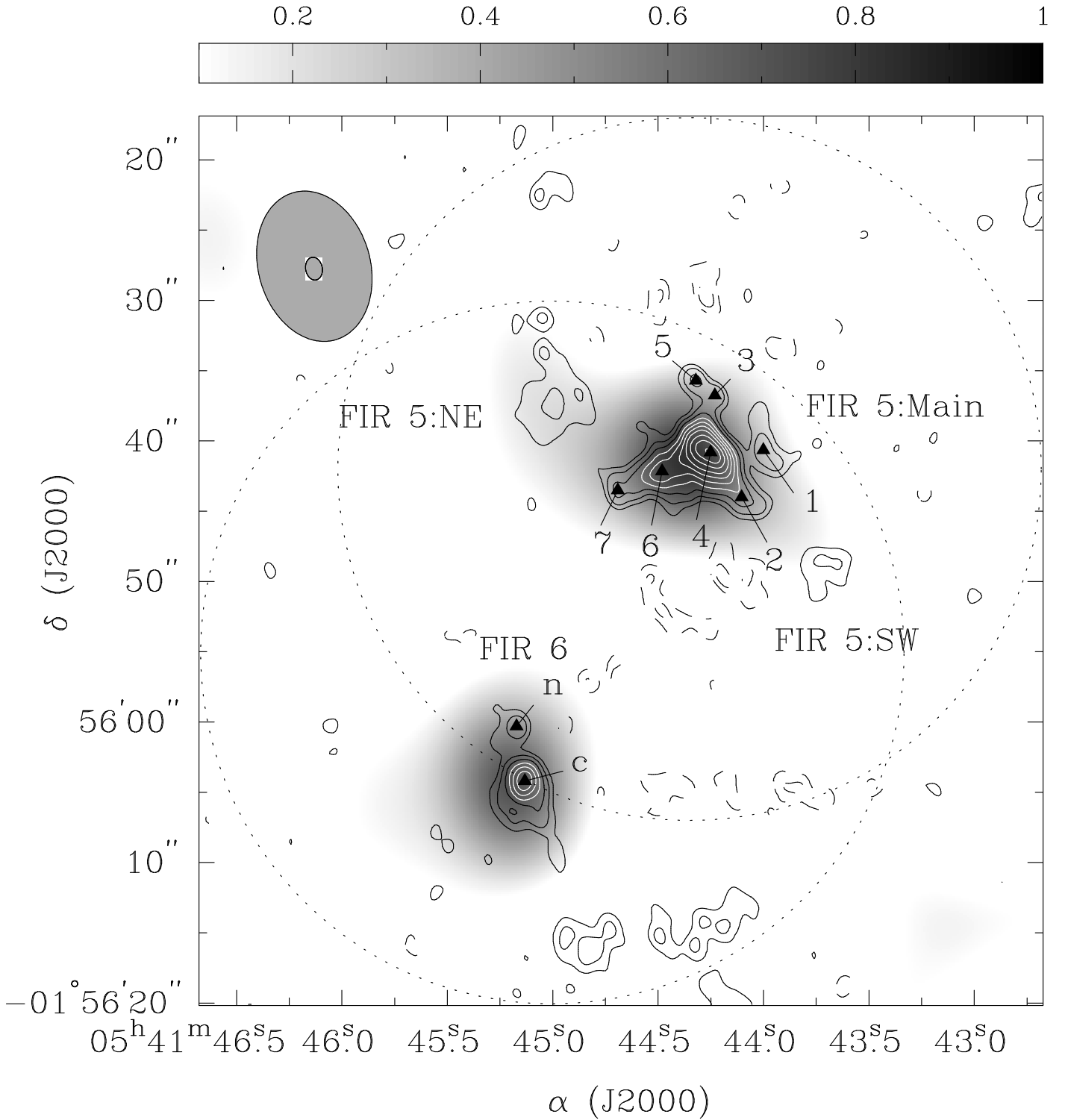}
\figcaption{Stokes $I$ maps of NGC\,2024 FIR\,5 and FIR\,6.
The contours show the total intensity (Stokes $I$) in the B+C-array data
at $-5, -3, 3, 5, 8, 12,17,23,30,40,50,60~\sigma$ levels
(1$\sigma$ = 3.3 \mJb).  The color of the contours is chosen to show 
better contrast on the background.
The field of view is represented by the northern dotted circle.   
The grey-scale shows the D-array data
and the field of view is the southern dotted circle. 
The grey ellipses in the upper left corner are the synthesized
beams for these two sets of data, which are 1\farcs6\x1\farcs2 with 
PA=11\deg\ for the B+C-array map and 11\arcsec\x8\arcsec\ with PA=16\deg\ 
for the D-array map. The triangles mark the positions of compact 
continuum sources identified in the B+C-array map (Table 1).}
\end{figure}

\begin{figure}
\plotone{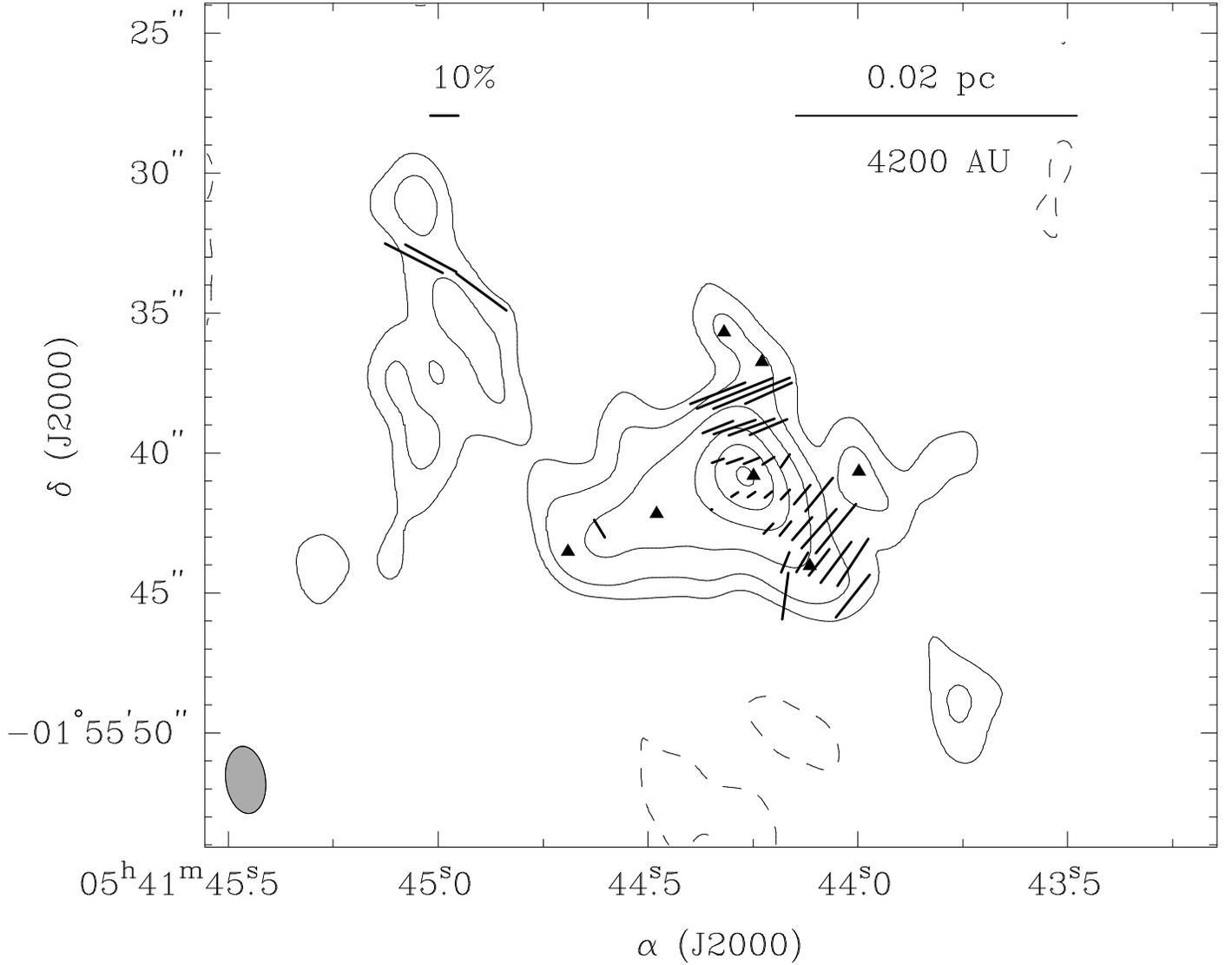}
\figcaption{Polarization map of NGC\,2024 FIR\,5.
The contours represent the mosaiced Stokes $I$ map 
at $-5, -3, 3, 5, 10, 20, 30, 40 \sigma$ levels.
The 1 $\sigma$ noise level of Stokes $I$ is 7.3 \mJb\ and
the beam HPBW is 2\farcs3\x1\farcs4 with PA=9\deg.
The line segments are polarization vectors (E-vectors), and
their lengths are proportional to the polarization percentage
with a scale of 10\% per arcsec length.
The triangles mark the positions of compact continuum sources
listed in Table 1.}
\end{figure}

\begin{figure}
\epsscale{0.5}
\plotone{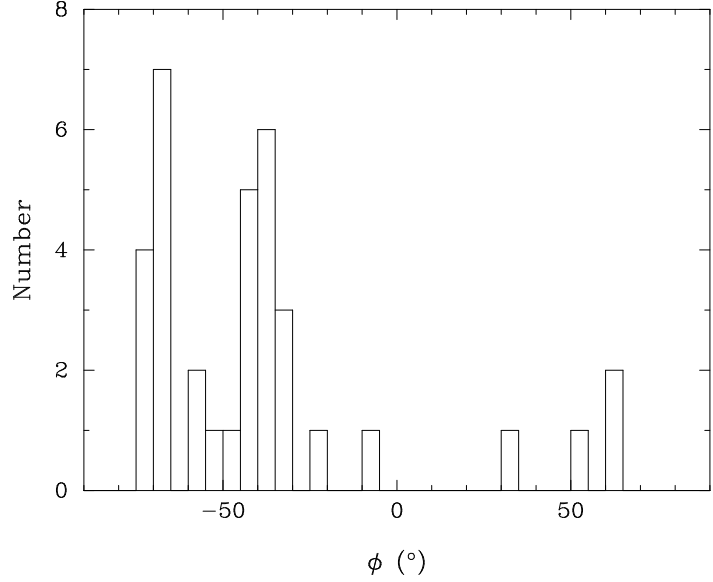}
\figcaption{The distribution of the position angles of the polarization
vectors in NGC\,2024 FIR\,5. The vertical axis is the total
number of the measurements in the bin.
}
\epsscale{1}
\end{figure}

\begin{figure}
\epsscale{0.5}
\plotone{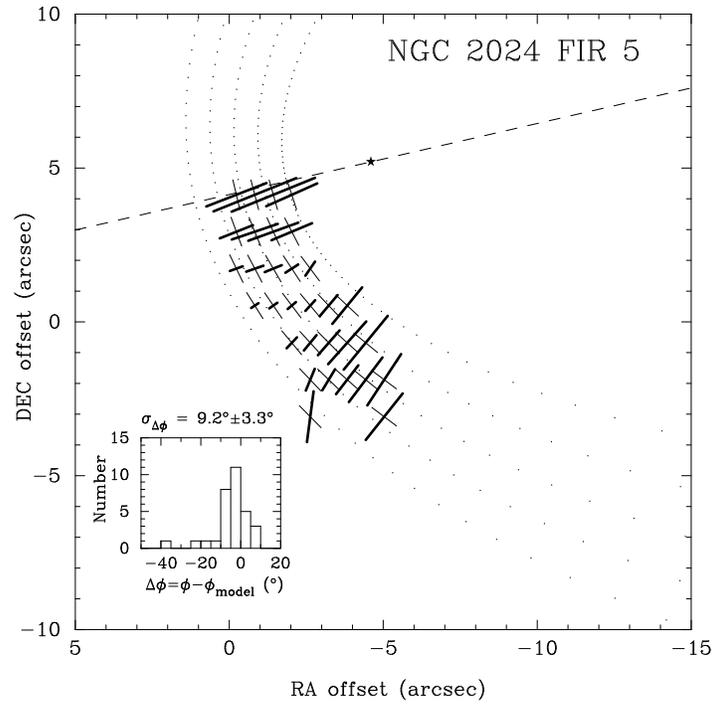}
\figcaption{Model fitting of the magnetic field morphology using a set of 
parabolas with the same focus point.  The line segments are polarization
vectors which are perpendicular to the magnetic field directions. 
The dotted lines shows the least-squares model, 
the dashed line is the symmetry axis, 
and the star is the common focus point.  
The histogram shows the distribution of the deviation between 
the field directions and the model.}
\epsscale{1}
\end{figure}

\begin{figure}
\epsscale{0.7}
\plotone{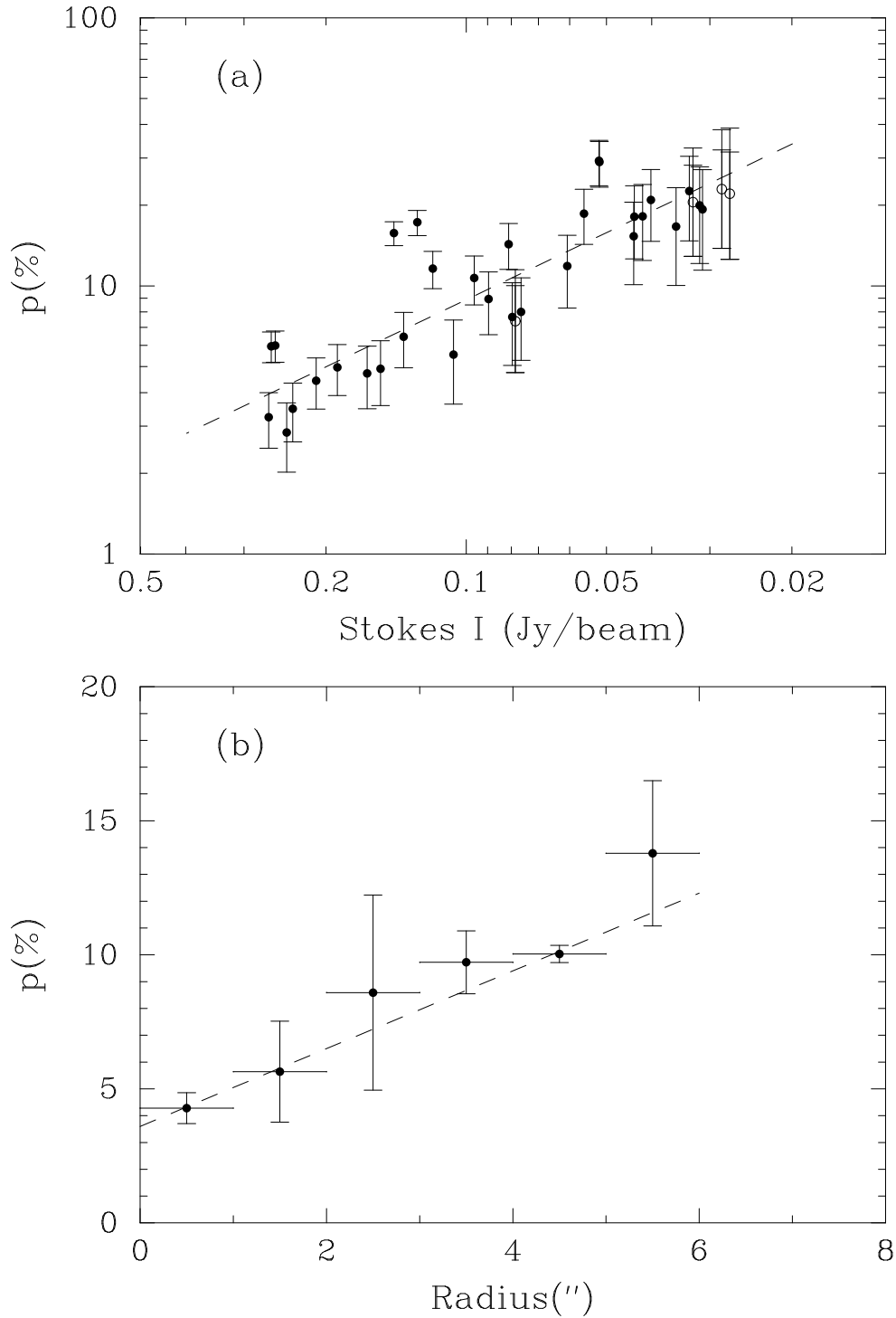}
\figcaption{Polarization percentage distribution.
The filled dots represent the data associated with the main core of FIR\,5, 
and the open dots represent the data associated with the eastern side of FIR\,5.
The dotted lines are the best least-squares fits for 
the set of data.
(a) shows the polarization percentage vs.\ the total intensity.
The error bar of each data point is the measurement uncertainty.
We plot the total intensity increasing to the left to emphasize
that high intensity corresponds to short distance to the center
of the core.
(b) shows the polarization percentage vs.\ the distance 
from the FIR\,5:LCGR 4 peak.  Data with R greater than 6\arcsec\
are not shown, because the average polarized intensity is lower 
than 1.5$\sigma_{I_p}$. 
The error bar in radius shows the range over which the data are
averaged and the error bar in $p$ is the standard deviation of the data 
in the range.  }
\end{figure}

\end{document}